\documentclass[11pt]{article}

\usepackage{graphicx}

\usepackage{epsfig}

\title{The infinite partition of a line segment  and multifractal objects}

\author{ A. I. L. de Ara\'ujo$^1$, R. F. Soares$^2$, J. P. de Oliveira$^3$, and  G. Corso$^{1,3}$  \\
     {$^1$ \small Departamento de F{\'{\i}}sica Te\'orica e Experimental,} \\
  { \small Universidade Federal do Rio Grande do Norte,} \\
  { \small Campus Universit\'ario  59078 970, Natal, RN, Brazil.}   \\
     {$^2$\small Departamento de Matem\'atica ,} \\
  { \small Universidade Federal do Rio Grande do Norte,} \\
  { \small Campus Universit\'ario  59078 970, Natal, RN, Brazil.} \\
     {$^3$ \small Departamento de Biof{\'{\i}}sica e Farmacologia,} \\
  { \small Universidade Federal do Rio Grande do Norte,} \\
  { \small Campus Universit\'ario  59078 972, Natal, RN, Brazil.} }
\date{}

\begin{document}

\maketitle

\begin{abstract}
We report an algorithm for the partition of a line segment according to 
a given ratio $\nu$. At each step  the length distribution among 
sets of the partition follows a binomial distribution. 
We call $k$-set to the set of elements with the same length at the step $n$. The total 
number of elements is $2^n$ and the number of elements in a same $k$-set is $C_n^k$. 
In the limit of an infinite partion this object become a multifractal where 
each $k$-set originate  a fractal. 
 We find the fractal spectrum $D_k$ and calculate where is  its 
maximum. Finally we find the values of $D_k$ for the
  limits $k/n \rightarrow 0$ and $1$.

\end{abstract}


{keywords: multifractal, binomial distribution, partition of a segment, 
spectrum of fractal dimensions. }

\section{Introduction}

Multifractals have been largely employed in the characterization of time series. This tool 
have been successfully applied in many different areas as economics \cite{economics}, meteorology \cite{meteor}, 
geology \cite{geo}, or biomedical \cite{biom}. Several algorithms have been used to find the multifractal spectrum 
of the time-series, for instance, wavelet analysis \cite{wavelets} 
and DFA (Detrended fluctuation analysis) \cite{dfa}.  However, despite 
the large use of multifractals as a time-series analysis technique there is no  
simple geometrical examples of multifractal sets.

Some years ago it was introduced  a multifractal partition of the unit 
square \cite{oum}. This mathematical object was originally 
developed to model multifractal heterogeneity 
in oil reservoirs and to study percolation on complex lattices. Afterwards, a 
generalization of this object to random partitions was performed to improve 
the model \cite{alea1,alea2}. In this work we explore the unidimensional 
version of this model. Indeed, the multifractal partition of the square, 
cube or hypercube are, in essence, derived from a multifractal partition of a 
line segment.

 Cantor sets, Peano curves or Sierpinski models  are very useful subjects 
to grasp the fundamentals of fractals. On the other side, the study of 
multifractal sets lack simple geometrical examples. 
This paper intend to fulfill this vacancy in the literature.  
The work is organized as follows. In section $2$ we introduce the partition 
of a line segment that produces a multifractal set. We focus our attention
 on the case of a constant  cutting ratio $\nu$. The 
binomial  distribution naturally arises in the construction of the 
partition algorithm. In section $3$ we analytically derive the multifractal 
spectrum $D_k$ of the partition  and find 
its limit cases $k/n \rightarrow 0$ and $1$. We use a method 
 similar to the boxcounting algorithm. 
Finally in section $4$ we give our   final remarks.

\section{The partition of a line segment according to a ratio}

In this section we introduce a partition of a line segment. We expose the method 
of construction of the partition as a recursive algorithm. We call the 
sets at step $n$ as father sets respect to  son-sets at step $n+1$. Each father 
set always give origin to two son sets. The sum of lengths of the son sets is 
equal to the length of the father set, that means, the dynamic rule of the 
algorithm is length preserving. 
 At each step of the algorithm we use a ratio $\nu$ to perform the partition of the 
father sets. 
In the next paragraphs we will detail the initial steps of the 
algorithm, but the multifractal, that shall be  shown in the next section, 
is properly defined in the limit of $n$ going to infinity.

\begin{figure}[ht]
\begin{center}
\resizebox{90mm}{!}{ \includegraphics[angle=0]{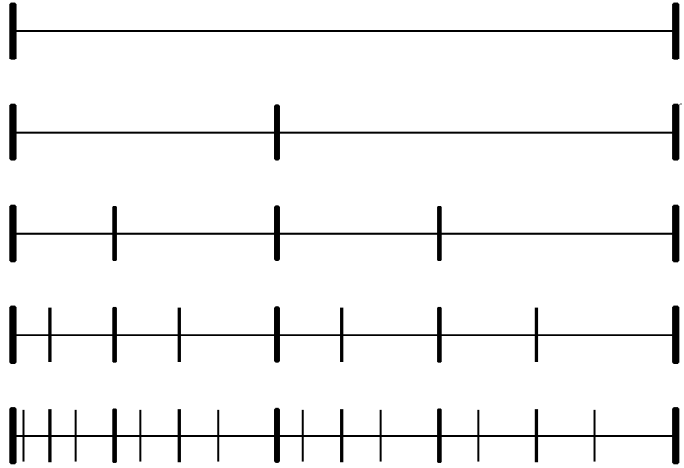}}
\caption{ A pictorial representation of the five initial steps of the partition 
of a line segment for a constant ratio. The partitions are indicated by vertical thick marks. For sake 
of clarity, the thick marks of the father-sets are always 
thicker than the son-sets. At  step $n$ the number of sets is $N_n = 2^n$. 
The elements with the same length form a $k$-set, there are $C_n^k$ elements in 
a $k$-set. The length distribution among  sets obeys a binomial rule.}
\label{fig1}
\end{center}
\end{figure}

In figure \ref{fig1} we show a representation of the five initial steps 
in the construction of our partition. The step $n=0$ corresponds to the 
segment of length $L$ itself. At step $n=1$ the segment is partitioned 
into two pieces of length $\nu L$ and $(1-\nu) L$. At step $n=2$ the two 
pieces are each one partitioned into two new sets giving origin to $4$ 
sets and so on. In table \ref{tab1} we show the iteration step $n$, the number 
of sets for each step $N_n = 2^n$ and the  partition length
(as a fraction of $L$) inside each step. We notice 
that the length distribution is trivially done by a binomial rule,  
that means, the partition gives raise, at step $n$, to a
binomial length distribution: 
\begin{equation}
    L = \sum _{k = 0}^n C_n^k  \nu^{k} (1-\nu)^{(1-k)} L
\label{bindidist}
\end{equation}

\begin{table}[ht]
\begin{center}
\caption{\footnotesize The main quantities of the algorithm of the partition of 
a segment line according 
to a ratio $\nu$. The iteration step $n$, the total number of 
sets at each step $N_n$, the full length partition as a fraction of $L$.}
\label{tab1}
\begin{tabular}{c|c|c}
\hline $n$ & $N_n$ & Partition length (in unities of $L$) 
\\ \hline
$0$ & 1 & 1
\\ \hline
$1$ & 2 & $\nu$ + $(1-\nu)$  
\\ \hline 
$2$ & 4 & $\nu^2 +2\nu (1-\nu) + (1-\nu)^2$
\\ \hline
$3$ & 8 & $\nu^3 +3\nu^2 (1-\nu) + 3\nu (1-\nu)^2 + (1-\nu)^3$
\\ \hline
 &  &  ...
\\ \hline
$n$ & $2^n$ & $ \sum _{k = 0}^n C_n^k  \nu^{k} (1-\nu)^{(1-k)} $
\\ \hline
\end{tabular}
\end{center}
\end{table}

We remark some  differences between the exposed process and the algorithm 
construction of Cantor set and Koch curve  \cite{fractal}. 
Initially the cited fractals do not conserve length along the dynamic formation, the 
measure divergence in these sets is originated in the algorithm process itself, the Cantor set 
subtracts subsets and the Koch adds subsets at each step. In addition,  in these fractal 
 models all sets, at the same step of their generating algorithm, 
have the same length. By that reason these models produce monofractal sets. 

We call  $k$-sets the set of all elements, at the step $n$, 
that has the same length $ \nu^{k} (1-\nu)^{(1-k)} L$. 
By construction there are $C_n^k$ elements in a $k$-set. In the next section, for the 
limit of $ n \rightarrow \infty$ we 
will determine the fractal dimension of each $k$-set and  characterize the 
set of all $k$-set as a multifractal object.

To find the length partition of a line segment in the 
 case of variable $\nu$ we proceed as follows. 
Consider that $\nu_i$ is present $n_i$  times at the partition and that at step $n$  we have 
  $n = n_1+n_2+...+n_l$. In this case the 
length distribution  is done  by:
\begin{equation}
    L = \frac{n!}{n_1! n_2! ... n_l!} \nu_1^{n_1} \nu_2^{n_2} ... \nu_l^{n_l} L
\label{multidist}
\end{equation}

In figure \ref{fig1} we assume a fixed rule in the partition. Each time a 
 father set is cut into two new pieces the largest 
 son set  is always situated at the right position. 
We notice that this is an arbitrary choice that will not change the 
length distribution of the object. On the other side, the topology of the partition (a 
subject that we do  not consider in this paper) will change since 
the neighborhood properties will change. We cite that the bidimensional version of the 
multifractal has a topological treatment \cite{topology} that can not be easily  extended 
to the unidimensional object. In fact, the topology of a line segment is 
sort of trivial since each open set in the segment 
 has always two neighbours. In the cited work \cite{topology} 
we report that the bidimensional multifractal version of the object we work in this 
paper has a power-law distribution of neighbours.

\section{The  multifractal spectrum}

In this section we estimate the fractal dimension for each $k$-set that 
results from the partition of a segment. 
We calculate the fractal dimension of each $k$-set, $D_k$, with help of the 
definition: 
\begin{equation}
   D_k = lim_{\epsilon \rightarrow 0} \frac{ log (N(k) ) }{log (1/\epsilon ) }
\label{DNN}
\end{equation}
In this definition $N(k)$ is the number of open balls, of size $\epsilon$, 
 necessary to cover a given $k$-set. 

In order to use (\ref{DNN}) we generate a proper coverage of the segment $L$.
To create an adequate coverage of the partition 
we take  at step one $L = s + r$ for $r,s$ integers. Therefore  
 $\nu$ is a rational number $\nu= r /(s+r)$ (we take $r < s$). 
In addition, at each step $n$  the size of the segment is $L=(s+r)^n$ and 
 the size of the open balls used to cover the $k$-sets is: 
$$        \epsilon_n = \frac{1}{L} = \frac{1}{ (s+r)^n }    $$

In this way, $N(k)$, the total length of each $k$-set is $C_n^k r^k s^{(n-k)}$ and as 
a result $D_k$ is done by: 
\begin{equation}
   D_k = lim_{n \rightarrow \infty} \frac{ log ( C_n^k \> r^k \> s^{(n-k)} ) }{ log  ((r+s)^n )}
\label{Dk}
\end{equation}

\begin{figure}[ht]
\begin{center}
\resizebox{100mm}{!}{ \includegraphics[angle=270]{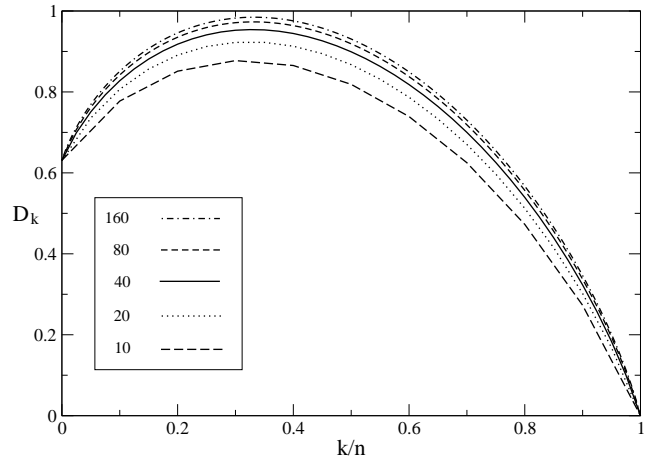}}
\caption{ The multifractal spectrum $D_k$ versus $k/n$.  The number of steps in the 
algorithm is $n=10,20,40,80$ and $160$ as indicated in the figure. 
We use $r=1$ and $s=2$  as a case study. }
\label{fig2}
\end{center}
\end{figure}

In figure \ref{fig2} we computationally test the finite size effect over the 
 multifractal spectrum (\ref{Dk}). We use as a case study  $r=1$ and $s=2$
and as a consequence $\nu=\frac{1}{2+1}=\frac{1}{3}$. In the simulations we 
take  $n=10,20,40$, $80$ and $160$ to follow the 
convergence of the spectrum. In this figure we plot $k/n$ in the horizontal 
axis to compare the results of several $n$. We observe in the data that 
in the limit of large $n$ the spectrum seems to touch the line $D=1$ which suggests that 
the  $k$-set of  largest dimension is dense in the line. 

It is interesting to estimate $D_k$ in the limits of $k$ going to $0$ and $n$. Using
equation (\ref{Dk}) we calculate $D_k$ for $k \rightarrow 0$:
$$
  D_k = \frac{ log ( C_n^0 \> r^0 \> s^{(n-0)} ) }{ log  ((r+s)^n )} = 
    \frac{ n \> log (1 s) }{n \> log (r+s)} = \frac{log(s)}{log(s+r)}.
$$
Summarizing,  we have: 
\begin{equation}
 k \rightarrow 0   \hspace{1cm} D_0 = \frac{log(s)}{log(r+s)}.   
\label{eq5}
\end{equation}
In a similar way we estimate:
\begin{equation}
   k \rightarrow n \hspace{1cm} D_n = \frac{log(r)}{log(r+s)}.
\label{eq6}
\end{equation}

We use the notation $ k \rightarrow 0$ instead of $k=0$ because 
he multifractal is only defined in the 
limit of $n \rightarrow \infty$, and therefore  $k=0$ is in fact an accumulation point. It is 
worth to note that in the limit of $ k \rightarrow 0$ or $ k \rightarrow \infty$ the fractal dimension is 
not necessarily $0$. One could think that $D_{k}=0$ for $k=0$ 
because the multifractal of the corresponding $k$-set for $k=0$ 
is composed by just one point and the dimension of a 
single point is zero. This argument is not valid 
because the multifractal is only properly defined in the 
limit of $ n \rightarrow \infty$, where the variable 
$k$ is not discrete, but continuous. 

\begin{figure}[ht]
\begin{center}
\resizebox{100mm}{!}{ \includegraphics[angle=270]{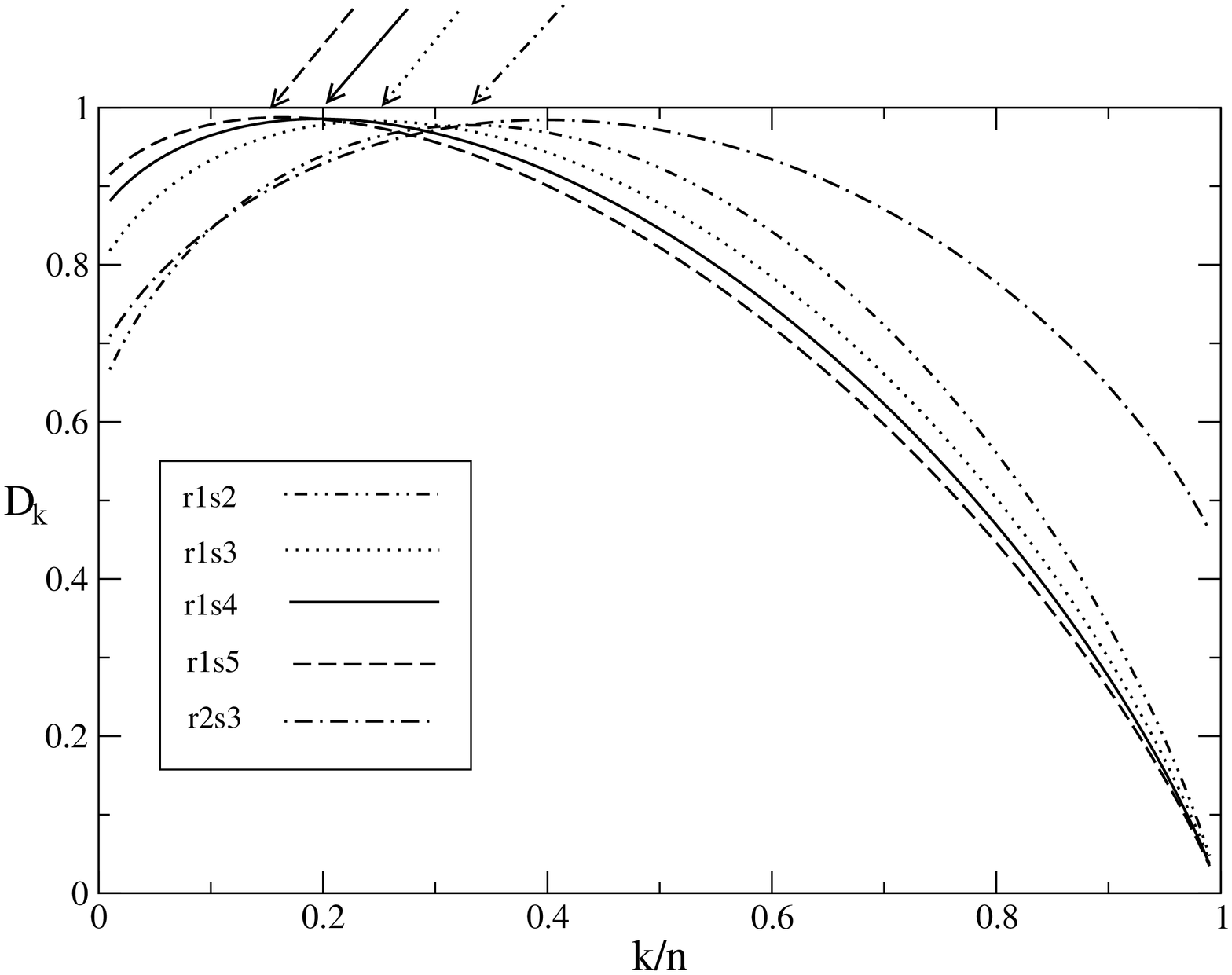}}
\caption{ The multifractal spectrum $D_k$  for 
several ratio partitions  as indicated in the figure. The maximum of 
each curve is evaluated in the text and indicated in the figure 
by arrows at the top of the figure. 
 The limits of $k/n \rightarrow  0$ and $1$ are analytically determined in the text. }
\label{fig3}
\end{center}
\end{figure}

We show the spectrum of fractal dimension for several ratios $\nu$ in 
figure \ref{fig3}. We display in this plot the spectrum $D_k$ versus $k/n$ for 
some $s$ and $r$ as indicated in the figure. In this estimation we use 
$n=100$. The relations (\ref{eq5}) (\ref{eq6}) are verified in the curves of the figure. 

We find the maximum of $D_k$ for any $\nu$ using the property of the maximum of the binomial 
distribution \cite{Meyer}. In this way the maximum of $D_k$ is at $\nu n$ or $\frac{r}{s+r} n$. To 
compare this result with the maximum of curves in figure \ref{fig3} we use the 
normalization $k/n$ and the maximum is at $\nu$. We check in the figure, for instance, that the 
maxima for the cases $r=1$ and $s=2,3,4$ and $5$ are at positions 
$\nu=\frac{1}{3},\frac{1}{4},\frac{1}{5}$ 
and $\frac{1}{6}$ respectively. These values are indicated by successive arrows in the plot.

The degenerate case $r=s=1$ corresponds to the 
standard partition of a segment which generates the unidimensional lattice. 
In this limit all elements of the partition have the same size.  
Following this rule we cover the line segment with $2^n$ equal elements of length 
$\epsilon = 1/2^n$ and the dimension of the object is not a fractal, 
indeed $ \frac{log 2^n}{log 2^n}= 1$ 
In this case there is just one  $k$-set   that  trivially have the same dimension of 
the line segment itself.

\section{Final remarks}
In this work we introduce an object that arises from an infinite partition 
of a line segment. We present the construction algorithm of the object as  
a recursive sequence.  In the case of a constant partition ratio $\nu$ the length 
distribution of its elements, at step $n$ of the algorithm, follows a binomial 
distribution.  At step $n$, the partition is formed by $2^n$ sets grouped in 
$(n+1)$ $k$-sets, that means, sets whose elements 
share the same length $ \nu^k (1-\nu)^{(1-k)}$. 
 In the limit of $n \rightarrow \infty$ 
each $k$-set is a monofractal set. The  multifractal set is composed by the totality 
of infinite monofractal sets. 
In addition we find the fractal spectrum $D_k$ and estimate where is  its maximum. 
 Finally we find the values of $D_k$ for the limits $k/n \rightarrow 0$ and $1$.

In this work we adapted to a line segment an algorithm of generation of multifractal objects that 
was initially defined on a square. 
The route from two dimensions to one dimension is a challenge. 
At one side we loose the attractive idea of modeling bidimensional 
geological formations. Otherwise, once we go 
to an unidimensional version of the problem some points became more clear:    
the generalization of this object to any dimension, 
its formation algorithm and the spectrum of fractal dimensions.

We think we have also attained our aims by showing to a general reader an illustrative 
 ans simple geometrical multifractal set. 
However, some  questions about this multifractal set remain open.
 Is the largest set of the multifractal dense in the line? The simulations
indicate a positive answer, but we lack a rigorous proof of that. Following equation (\ref{multidist}), is there 
a simple spectrum of dimensions for the case of non constant partition? 
 The answer seems positive if we could define 
appropriate $k$-sets. A more general question would be: 
if the set of ratios $\nu_i$ in equation (\ref{multidist}) follows a normal distribution, 
what will be the resulting multifractal?
 We conjecture that this new object would be usefull  as a
 benchmark to compare with empirical multifractals.

\vspace{0.8cm}

\centerline{\bf Acknowledgments}

We thank the financial support from  CNPq (Conselho Nacional de Pesquisa).


%

\end{document}